\input{aipcheck}

\documentclass[
    ,final            
  ]
  {aipproc}

\layoutstyle{8x11single}

\begin{document}

\title{Dark Matter Seeding in Neutron Stars}


\author{M. \'Angeles P\'erez-Garc\'ia}{
  address={Department of Fundamental Physics and IUFFyM, \\Universidad de Salamanca, Plaza de la Merced s/n 37008 Salamanca}
}

\author{Joseph Silk}{
  address={Oxford Physics, University of Oxford, Keble Road OX1 3RH, Oxford, United Kingdom.}
}

\author{Jirina R. Stone}{
  address={Oxford Physics, University of Oxford, Keble Road OX1 3RH, Oxford, United Kingdom and\\ Department of Physics and Astronomy, University of Tennessee, Knoxville TN 37996, USA}
}

\begin{abstract}
We present a mechanism that may seed compact stellar objects with stable lumps of quark matter, or {\it strangelets}, through the self-annihilation of gravitationally accreted WIMPs. We show that dark matter particles with masses above a few GeV may provide enough energy in the nuclear medium for quark deconfinement and subsequent strangelet formation. If this happens this effect may then trigger a partial or full conversion of the star into a strange star. We set a new limit on the WIMP mass in the few-GeV range that seems to be consistent with recent indications in dark matter direct detection experiments.
\end{abstract}

\maketitle


\section{Accreting Dark Matter in neutron stars}

 There is compelling evidence \cite{rev} that as much as $90\%$ of the matter in the Universe is under the form of dark matter (DM). A set of very popular candidates for DM are WIMPs (weakly interacting massive particles). This mostly follows from what is called the {\it WIMP miracle} since cosmological data are consistent with a thermal relic cross section $<\sigma_{\rm annih} v> \approx 10^{\rm -26}\,cm^{\rm 3} s^{\rm -1}$. WIMPs from the galactic halo can be accreted onto compact massive objects such as neutron stars (NS) \cite{kouvaris2008} or white dwarfs.
The signature of WIMP self-annihilations (assuming WIMPs are Majorana particles) is energy release in the range equivalent to twice the WIMP mass, leading to subsequent particle production and/or heating of the star. Current predictions of the WIMP mass span the range from m$_\chi\approx$ 1 $\rm GeV/c^{\rm 2}$ up to 10 $\rm TeV/c^{\rm 2}$ however recent results from direct detection experiments DAMA and CoGeNT seem to favour a light DM particle in the range 4-12 $\rm GeV/c^{\rm 2}$ \cite{lightwimp}.
 
Compact objects such as black holes or NS are considered to be gravitational accretors of DM. NS configurations, yielding gravitational mass, $M$, and radius, $R$, are computed using an equation of state (EoS) that relates pressure to energy density and temperature, $p=p(\epsilon, T)$. A typical NS has a mass M$ \approx$ 1.4 M$_\odot$, radius $R \approx 10\,$ km, surface temperature $T_{\rm surf} \approx 100 \,$keV and central mass density $\rho \approx 1.0 \times$ 10$^{\rm 15}$ g/cm$^{\rm 3}$. There is a variety of model predictions for  NS  composition  \cite{lattimer2007} but there is as yet no consensus about the fundamental nature of the matter in NS interiors.

The accretion rate of WIMPs captured by a typical NS is given by 
\cite{press}
\begin{equation}
{\cal F}= \frac{3.042 \times 10^{25}}{ m_{\chi} (\rm GeV)} \frac{\rho_{DM}}{\rho_{DM,0}}\, (s^{-1}),
\label{rate}
\end{equation}
assuming a WIMP-nucleon interaction cross section $\sigma_{\chi n} > 10^{-45}$ and $\rho_{\rm DM,0}=0.3\, \rm GeV/cm^{\rm 3}$. Due to competing effects of annihilation and evaporation, the number of accreted WIMPS at time $t$ is obtained by solving the differential equation ${\dot N}= {\cal F}-\Gamma_{\rm annih}-\Gamma_{\rm evap}$ where $\Gamma_{\rm annih}$ is the self-annihilation rate calculated as $\Gamma_{\rm annih}=\langle\sigma_{\rm annih} v\rangle \int n^2_{\chi} dV=C_{\rm A}N(t)^{\rm 2}$, $\langle\sigma_{\rm annih}v\rangle$ is the product of thermally-averaged WIMP self-annihilation cross section and velocity, and  $n_\chi=\frac{\rho_{\rm DM}}{m_{\chi}}$, the number density of WIMPs inside the NS, is assumed to be constant. The evaporation rate, $\Gamma_{\rm evap}$, decays exponentially with temperature  $\sim e^{\rm -G M m_{\chi}/R T}$,  and is negligible with respect to the annihilation rate for NS  with internal T $\sim$ 0.1 MeV \cite{krauss1986}. With this simplification, the population of WIMPs at time $t$ is given by $$ N(t)= ({\cal F} \tau) {\rm tanh}(t/ \tau)$$ where the time-scale is $\tau=1/{\rm \sqrt{{\cal F }C_{\rm A}}}$. For $t>>\tau$, when the equilibrium between accretion and annihilation has been reached, the number of particles accreted is time-independent, $N={\mathcal F}\tau$.

Assuming a regime when velocities and positions of the WIMPs follow a Maxwell-Boltzmann distribution with respect to the centre of the NS, the thermalization volume in the compact star has a radius $r_{\rm th}= (\frac{9 k T_c}{4 \pi G \rho_c m_{\chi}})^{1/2}=64 \, (\frac{T}{10^{\rm 5} K})^{1/2} (\frac{10^{\rm 14} {\rm g/cm^3}}{\rho_{\rm c}})^{1/2} (\frac{100 \,{\rm GeV/c^{\rm 2}}}{m_{\chi}} )^{1/2}\,(\rm cm)$. 
Taking typical NS conditions as central internal temperature $T_c=10^{\rm 9}$ K, $\rho_{\rm c}$=10$^{\rm 15}$ \rm g/cm$^{\rm 3}$ and m$_{\chi}=1\, \rm GeV/c^2-10\, TeV/c^2$, r$_{\rm th}\approx 2\times 10^{5}-2 \times 10^{2}\,{\rm cm}$. Then the thermalization volume is  $V_{\rm th}=\frac{4}{3} \pi r_{\rm th}^{\rm 3} \approx  (10^{\rm -2}-10^{\rm -8})\, V_{\rm NS}$. 

\section{Dark Matter annihilation in the NS core}

The exact energy released due to WIMP self-annihilation is dependent on the nature of WIMPs and on the output  product channel. We parametrize this lack of knowledge by assuming a fraction of the annihilation energy is given by the efficiency factor, $f$, typically $f \approx 0.01-1$. The rate of energy released in annihilation processes in the star in the thermal equilibrium volume can be expressed as \cite{prl-silk},
\begin{equation}
{\dot E_{\rm annih}} = C_A N^2 m_{\chi}c^2 = f {\cal F} m_{\chi}c^2,
\label{edot}
\end{equation}
Taking $\rho_{DM}$=$\rho_{\rm DM,0}$ and an efficiency rate $f=0.9$, we have that in the range $m_{\chi} \approx 1-10^4$ $\rm GeV/c^{\rm 2}$ ${\dot E_{\rm annih}}=2.74 \times 10^{\rm 25}- 10^{\rm 29}\,{\rm GeV/s}.$ The energy released will convert partly into heat, which can stimulate $u,d$ quark bubble formation. Thermal fluctuations may arise in metastable hadronic matter via strong interactions \cite{olesen1994}. The timing and conditions of these transitions depend strongly on the pressure in the central regions of the star given by the EoS of hadronic matter. Then $ud$ matter undergoes non-leptonic weak reactions, such as $u+d \leftrightarrow u+s$, to form drops of strange $uds$ matter, which has lower energy as a result of the reduction in Fermi energies through the introduction of a new flavor. The stability, among other properties of these drops, commonly called {\it strangelets}, has been extensively studied (see e.g. \cite{sqm}) and depends on the electrical charge, strangeness fraction and size. 
These exotic forms of matter are currently being searched  with the CASTOR in the LHC CMS experiment in the forward rapidity and above earth attached to the ISS with the AMS spectrometer \cite{castor-ams}. The energy needed to form a stable long-lived strangelet of baryonic number $A$ can be calculated from its quark constituents, in a first approximation, using either the  MIT bag model with shell mode filling or the liquid drop model (for details see e.g. \cite{madsen2}). These calculations show that stability conditions point towards  large negative electrical charge fractions and high strangeness. In the ideal Fermi-gas approximation, the binding energy of a strangelet with baryonic number $A$ composed of massive quarks of flavor $i$ $(i=u,d,s)$ is
\begin{equation}
E^A(\mu_i, m_i, B) =  \sum_i (\Omega_i + N_i \mu_i)+BV 
\end{equation}
where $N_i $ is the quark number in the strangelet of baryon number $A = \frac{1}{3} (N_{\rm u}+N_{\rm d}+N_{\rm s})$, $\mu_i(n_{\rm A})$ is the quark chemical potential at baryonic number density $n_{\rm A}$. The thermodynamical potential for the $ith$-quark type with mass  $m_i$ is given by $\Omega_i (\mu_i, m_i)=\Omega_{i ,V}V+ \Omega_{i, S} S +\Omega_{i ,C} C$ being $V=A/n_A=\frac{4}{3} \pi R^3$ the volume, $S=4\pi R^2$ the surface and $C=8 \pi R$ the curvature of the spherical strangelet. Expressions for these potentials can be obtained from \cite{madsen2}. The masses of the quarks are taken as $m_u=2.55$ MeV, $m_d=5.04$ MeV, $m_s=104$ MeV respectively. In order to consider charged strangelets a Coulomb correction term can be added as $E_{\rm coul}=\frac{4}{3}(\frac{\alpha Z^2_V}{10R} +\frac{\alpha Z^2}{2R})$ with $Z_V=\displaystyle \sum_i q_i n_i V$ and $Z=\displaystyle \sum_i q_i$ given the $ith$-quark charge  and number density, $q_i$ and $n_i$. Then, the energy of a strangelet with baryon number $A$ is  $E^A_{\rm slet}\approx E^A(\mu_i, m_i, B)+E_{\rm coul}.$

Estimation of the minimum value of $A$ in a long-lived strangelet is model-dependent but typical minimum $A$ values are in the range $A_{\rm min} \approx 10-600$ \cite{alford2006}. Strangelets with `magic' numbers of quarks are found for bag constant values $145 \leq B^{1/4}  \leq 170$ MeV. If smaller strangelets than the minimum $A$ are created, they will decay rapidly, however the long-lived ones have lifetimes of days.
Since this time-scale is larger than the conversion time-scale $t_{\rm conv} \approx 100\,s $ \cite{bha2006}, it is in principle possible that strange stars could be formed if this conversion is triggered. As an estimate the rate of formation would be $\dot N_{\rm slet}=\dot E_{annih}/E^A_{\rm slet}.$
For $f=0.9$, $A=10$, $n_A=n_{A,0}$, $m_{\chi}=1$ $\rm GeV/c^{\rm 2}$, a number ${\dot N_{\rm slet}} \approx 10^{\rm 23}  \,s^{-1}$ are created. Assuming that in the center of the star,  self-annihilations are dominant, the energy deposited in the medium, per annihilation,  will be $2 f m_{\chi}c^2$. This allows us  to set a limit on the energy scale, at a given central baryonic density $n_A$. 
\begin{equation}
2 f m_{\chi}c^2 \geq E^A_{\rm slet}(\mu_i (n_A), m_i, B).
\label{ineq}
\end{equation}

We now discuss the results, a more detailed explanation can be found in \cite{prl-silk}. We have found that WIMP mass shows a weak dependence on nuclear physics input, parameterized by $B$. As the central baryon density increases, the lower limit on the WIMP mass decreases below  $\sim \rm 4\, GeV/c^2$. We adopt  $n_A=5 n_{A,0}$ as the highest value of the central baryon density, although recent pulsar masses constrain central densities to somewhat smaller values.

\begin{figure}[hbtp]
\includegraphics [angle=0,scale=1.1] {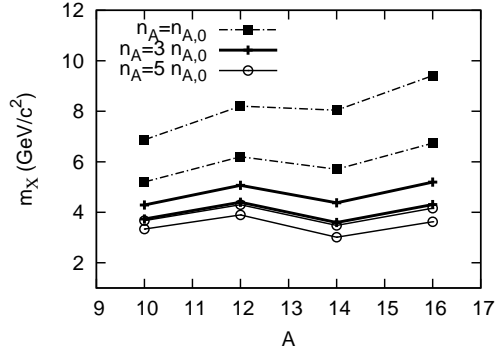}
\caption{ WIMP mass as a function of long-lived strangelet baryon number $A$ for  limiting values of $B^{1/4}=140, 170$ MeV, and $n_A/n_{A,0}=1,3,5$.}
\label{Fig1}
\end{figure}

In Fig. \ref{Fig1}, we show the WIMP mass to produce a long-lived strangelet as a function of $A$ for an efficiency rate $f=0.9$ and different values of the central density $n_A=n_{A,0}$
(dash-dotted line), $3 n_{A,0}$ (thick solid line), $5 n_{A,0}$  (thin solid line)  for limiting values of the bag constant. For each  pair of curves, the lower line is for $B^{1/4}=145$ MeV and the  upper line is for $B^{1/4}=170$ MeV. For a typical baryonic density of $\sim 3 n_{A,0}$, a lower limit for the WIMP mass $m_{\chi}\geq 4$ $\rm GeV/c^{\rm 2}$ is predicted.

Since the mass of the WIMP sets a scale of the possible triggering of the strangelet formation we find that as more massive WIMPs annihilate, the smaller the efficiency rate  becomes that is required to trigger a conversion from nucleon to strange quark matter. 
In summary, through the mechanism of self-annihilation of DM candidates in the central regions of a typical NS proposed in this work, we have derived a limit on the mass of DM particle candidate using the fact that there is a minimum long-lived strangelet mass needed to trigger a conversion from nuclear matter into strange quark matter. In the range $m_{\chi} \geq 4$ $\rm GeV/c^2$ the energy released in WIMP self-annihilation is sufficient to burn nucleon matter into a long-lived strangelet that may trigger full conversion to a strange star. This work presents a scenario compatible with current experimental direct and indirect DM searches and may have relevant astrophysical implications.


\begin{theacknowledgments}
M.A.P.G. would like to thank the University of Oxford for kind hospitality. M. A. P. G. and J. R. S. are members of the COMPSTAR collaboration. M. A. P. G. acknowledges partial support under MICINN projects MULTIDARK consolider, FIS-2009-07238  and Junta de Castilla y Le\'on Excellence group GR-234.
\end{theacknowledgments}



\bibliographystyle{aipproc}   


\end{document}